**Evaluating Lightweight Block Cipher Payload Encryption for Real-Time CAN Traffic**

Kevin Setterstrom
Department of Computer Science
North Dakota State University
1320 Albrecht Blvd., Room 258
Fargo, ND 58108
P: +1 (701) 231-8562
F: +1 (701) 231-8255
E: kevin.setterstrom@ndsu.edu

Jeremy Straub[1]
Center for Cybersecurity and AI
University of West Florida
220 W. Garden St., Suite 250
Pensacola, FL 32502
P: +1 (850) 474-2999
E: jstraub@uwf.edu

**Abstract**

This study evaluates the feasibility of integrating lightweight block cipher payload encryption into a real-time embedded controller area network (CAN) node using a QT PY ESP32-S2 microcontroller. This work seeks to determine whether the use of a block cipher can prevent semantic taxonomy-based reverse engineering, which infers signal meaning from unencrypted CAN traffic using observation and statistical analysis. CAN payloads are encrypted using a lightweight block cipher and evaluated through experiments that measure timing impact, payload pattern observability, and correlation-based inference. Results indicate that encryption masks constant values and predictable signal patterns while preserving a 100 Hz transmission schedule. These findings suggest that lightweight payload encryption can reduce passive, observation-based inference of CAN signal semantics on resource-constrained hardware with limited timing overhead impact.

**Keywords:** reverse engineering, controller area network, CAN bus, cybersecurity, cryptography

## 1. Introduction

The security of serial networks in automotive electronic control units (ECUs) has become a significant concern following several high-profile vehicle hacking incidents that exposed fundamental vulnerabilities in in-vehicle communication systems [1], [2], [3]. Central to these networks is the controller area network (CAN), which, despite its widespread adoption and critical role in automotive communication, provides no native support for encryption according to the CAN 2.0 specification [4]. As a result, CAN traffic is vulnerable to a range of attacks, including semantic taxonomy-based reverse engineering techniques that infer signal meaning through statistical correlation and observation [5], [6]. These vulnerabilities limit the effectiveness of security-by-obscurity approaches traditionally relied upon in automotive systems.

---
[1] This work was partially conducted while J. Straub was at North Dakota State University.

The absence of encryption in CAN networks is largely due to the historical difficulty of deploying cryptographic mechanisms on resource-constrained embedded platforms [7]. Many ECUs operate under strict real-time and memory constraints, making conventional cryptographic algorithms impractical. Lightweight block cipher implementations, such as Speck [8], are designed to provide security within the computational, memory, and power constraints of embedded environments [9], [10].

In this work, the SpeckSmall encryption library [11] is implemented on a QT PY ESP32-S2 microcontroller to evaluate the feasibility of encrypting CAN payload data for real-time automotive communications. The objectives are to assess whether lightweight encryption can mitigate semantic taxonomy-based reverse engineering attacks and preserve the timing characteristics required by real-time CAN systems. Through experimental evaluation, this study examines the impact of Speck encryption on message timing, data pattern visibility, and signal correlation. The results indicate that lightweight payload encryption can be integrated into an embedded CAN node while preserving timing behavior at 100 Hz.

This study focuses on passive, observation-based inference of CAN signal semantics. Encryption may also provide protection against other attacks including message injection, spoofing, denial of service, and key compromise. The results related to encryption's timing impact may aid future efforts in these areas which are outside the scope of this work.

This paper continues, in Section 2, with a review of prior foundational work. Then, Section 3 presents the methodology used in this study. Section 4 presents the results of the study and Section 5 discusses their implications. The paper then concludes, in Section 6, and discusses areas of potential future work.

## 2. Background

This section provides a overview of CAN, followed by a discussion of CAN security considerations in automotive systems. Relevant prior work on CAN security and lightweight cryptography, which provides a foundation for the work in this paper, is reviewed.

### *2.1. Technical Overview of Controller Area Network*

The CAN protocol was introduced by Bosch in the early 1980s to reduce wiring complexity and support the growing number of electronically controlled automotive subsystems [12]. Since its introduction, CAN has become the dominant in-vehicle communication protocol, with early adoption by manufacturers such as Mercedes-Benz [13]. By enabling distributed electronic control, CAN replaced centralized architectures with networked subsystems capable of coordinated, real-time data exchange [12] .

One notable example of this is the transition to drive-by-wire technology, where mechanical linkages are replaced by electronically mediated control [14]. In throttle-by-wire systems, pedal position sensors digitize driver input and transmit this information to an engine control module, which commands an actuator to regulate throttle position [14], [15]. These components exchange electronic control signals in real time, enabling distributed coordination between sensors, controllers, and actuators. This architecture provides greater flexibility, integration capability, and improved system-level performance, as compared to mechanical implementations [14].

Modern vehicles rely on numerous ECUs for functions ranging from driver assistance to connectivity and powertrain control [16]. using in-vehicle network protocols, such as CAN, LIN, and FlexRay. CAN is the industry standard for vehicle networks [16].

Figure 1 illustrates the contrast between a control network implemented using direct point-to-point wiring and one implemented using CAN. Without CAN, each device requires dedicated connections to all other devices that it must communicate with. This results in increased wiring complexity. With CAN, a shared communication bus enables the same functional interactions with substantially less wiring required.

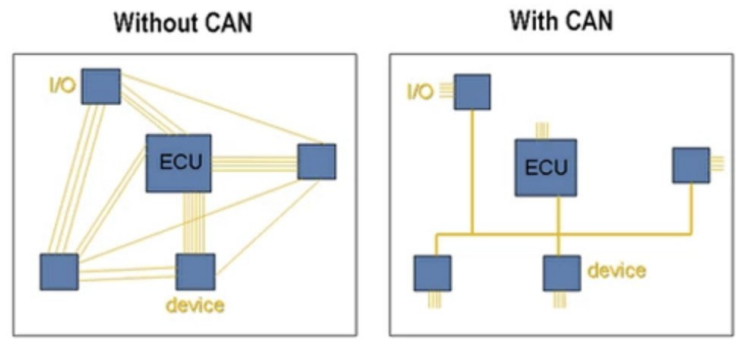

**Figure 1.** Automotive control network without CAN (left) vs with CAN (right) [17].

CAN communication is organized into structured message frames. Figures 2 and 3 illustrate the standard 11-bit and extended 29-bit CAN frame formats, respectively. They each show the identifier, control fields, data payload, and error-checking components as commonly implemented in CAN controllers [18].

| SOF | 11-bit Identifier | RTR | IDE | r0 | DLC | 0…8 Bytes Data | CRC | ACK | EOF | IFS |
|---|---|---|---|---|---|---|---|---|---|---|

**Figure 2.** Standard CAN: 11-Bit Identifier [18].

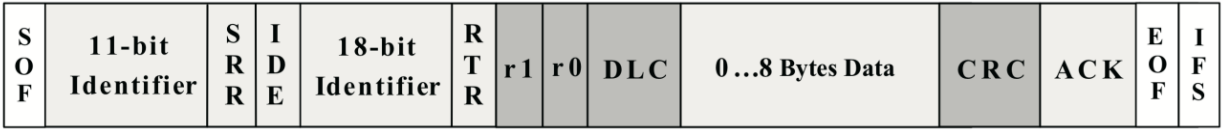

**Figure 3.** Extended CAN: 29-Bit Identifier [18].

## 2.2. Review of CAN Bus Security

The CAN protocol is widely used for communications in modern automotive systems; however, the absence of native security mechanisms exposes CAN traffic to a range of attacks. These include semantic taxonomy-based reverse engineering-based attacks. Alfardus, et al. [19] provide an analysis of CAN security vulnerabilities, examining attack surfaces, threat vectors, and proposed mitigation strategies. Their analysis shows that CAN lacks inherent support for confidentiality, authentication, and access control, leaving in-vehicle networks susceptible to both passive and active attacks that can impact vehicle safety, privacy, and system integrity.

Alfardus and Rawat [19] identified several recurring security deficiencies in CAN-based systems. There are confidentiality issues. CAN messages are transmitted in plaintext, enabling adversaries with bus access to observe and analyze message contents. There are also integrity issues. The protocol does not authenticate message sources, allowing malicious nodes to inject falsified messages onto the bus.

Availability issues also exist. CAN arbitration mechanisms can be exploited to monopolize the bus using high-priority messages, resulting in denial-of-service conditions. Attackers with physical access to vehicle interfaces can also directly observe or manipulate CAN traffic. Additionally, remote attacks are possible. Vulnerabilities in telematics or infotainment systems can provide indirect access to the CAN bus from external networks. Finally, there are privacy concerns. CAN traffic may expose driver behavior and vehicle state information that can be leveraged for tracking or profiling.

Among these concerns, the lack of confidentiality enables passive observation and statistical analysis of CAN traffic, forming the basis for semantic taxonomy-based reverse engineering attacks. This work focuses on mitigating this class of attacks by evaluating the feasibility of encrypting CAN payload data on resource-constrained embedded platforms.

## 2.3. Literature Review

Prior research has explored a range of cryptographic mechanisms to improve the security of CAN communications while respecting the protocol's real-time and resource constraints. Existing approaches can be broadly categorized into lightweight payload encryption techniques and authentication-focused schemes.

Jukl, et al. [9] evaluated the use of the Tiny Encryption Algorithm (TEA), a lightweight block cipher, to encrypt CAN messages in a real-time environment. Their results demonstrated that payload encryption could be performed within CAN timing constraints at a baud rate of 250 kbit/s. This suggests that lightweight cryptography could be feasible on constrained CAN platforms.

Other efforts have focused primarily on authentication and message integrity. Herrewege, et al. [20] introduced CANAuth, a backward-compatible authentication protocol based on hash-based message authentication codes (HMACs). It is designed to prevent message spoofing and replay attacks. Similarly, Ueda, et al. [21] proposed a centralized authentication and monitoring mechanism for in-vehicle CAN networks that combines message authentication codes with key exchange to mitigate unauthorized message injection.

Dariz, et al. [7] analyzed a combined encryption and authentication scheme for CAN that appends an integrity tag to the payload and encrypts the resulting block. They evaluated confidentiality, integrity, and authenticity of the payload under the real-time constraints of CAN networks. They observed that 64-bit block cipher constructions align naturally with the fixed CAN payload size, avoiding fragmentation. However, their analysis also highlighted safety trade-offs, particularly in terms of residual error probability (raising concerns about diminished error detection effectiveness, in particular) when encryption is introduced. This work demonstrates both the feasibility and the architectural implications of using symmetric encryption in embedded CAN environments.

Fassak, et al. [22] proposed an elliptic curve cryptography (ECC)-based protocol for ECU authentication and establishing session keys for message authentication. The established keys are used to generate MACs and strengthen resistance against spoofing and message manipulation without modifying the CAN frame format. Halabi, et al. [23] introduced a lightweight symmetric encryption scheme, inspired by blockchain-style hash chains, where encryption keys are regenerated dynamically after each transmission. Their approach provides confidentiality and integrity protection against sniffing, replay, and spoofing attacks while maintaining compatibility with standard CAN frame formats and resource-constrained embedded platforms

While these studies demonstrate a variety of effective security mechanisms for CAN networks, many focus on authentication protocols, key management, or combined security frameworks. In contrast, this work isolates the feasibility and system-level impact of lightweight payload encryption as a countermeasure against passive semantic taxonomy-based reverse engineering, with an emphasis on the devices' real-time performance and implementation simplicity on resource-constrained hardware.

### *2.4. Lightweight Cryptography and Block Ciphers*

Embedded systems that communicate over protocols like CAN often operate under strict constraints on their processing capability, memory footprint, and energy consumption. As a result, conventional cryptographic algorithms can be impractical for direct deployment in these environments, motivating the development of lightweight cryptographic primitives [10].

Block ciphers are a central building block in many lightweight cryptographic designs, due to their efficiency and suitability for fixed-size data. CAN messages, which contain short and bounded payloads, align naturally with block cipher operation, making symmetric encryption a practical choice for protecting CAN traffic under real-time constraints. Prior studies have shown that block cipher implementations can achieve acceptable latency while providing confidentiality and supporting message integrity mechanisms when required [7], [10].

Lightweight block ciphers are specifically designed to minimize computational overhead and memory usage while ensuring predictable execution time. This is an essential property for real-time embedded systems. These characteristics make them particularly well-suited for automotive control environments where encryption must not interfere with deterministic communication behavior.

The Speck cipher was designed with these constraints in mind. It targets platforms with limited resources while enabling efficient software-based implementations [8]. Its structure and performance characteristics have led to its use in experimental and embedded security applications, including prior work on CAN-based systems [7]. In this study, Speck is employed as a representative lightweight block cipher to evaluate the feasibility and system-level impact of payload encryption in a real-time CAN environment.

## 3. Methodology

This section describes the experimental methodology that is used to evaluate the impact of lightweight encryption on a real-time CAN communication system. First, the encryption strategy is presented. Then, the hardware and software implementations are discussed. Finally, the experimental design is described.

### *3.1. Encryption Strategy*

To mitigate the vulnerabilities associated with semantic taxonomy-based reverse engineering of CAN messages, the lightweight SpeckSmall cryptography library [11] was selected to perform payload encryption. During early evaluation, it was observed that encrypting CAN payloads without additional variability produced deterministic ciphertexts for identical message contents, leaving the system susceptible to replay and pattern-based analysis.

To address this limitation, a freshness value was introduced into each CAN payload. The freshness value consists of a single byte of randomly generated data that is prepended to the application payload prior to encryption. Incorporating this value ensures that identical application data produces different encrypted outputs for each transmission, reducing deterministic ciphertext repetition. Note that the freshness value alone does not prevent replay acceptance unless paired with a receiver-side check mechanism of some sort. In this proof-of-concept implementation, the freshness value occupies the first byte of the CAN data field, followed by application data, resulting in a payload structure that remains compliant with standard CAN frame formatting.

Within the CAN frame structure illustrated in Figures 2 and 3, only the data payload field (data bytes 0–8) was encrypted in this study. The identifier field of the CAN frame (11-bit standard or 29-bit extended) was left unmodified, as it determines message arbitration and priority on the bus. Altering or encrypting the identifier would disrupt the arbitration behavior and violate established protocol semantics and interface definitions [17].

Although encrypting the CAN data field conceals the payload contents, it does not eliminate metadata leakage. The arbitration identifier and transmission timing remain observable on the bus, enabling an adversary to perform traffic analysis. From these observable characteristics, an attacker could infer active message identifiers, transmission frequency, burst patterns, priority relationships, and ECU communication structure. While this metadata does not reveal signal semantics, bit-level structure, scaling factors, or plaintext values, it may still expose higher-level system characteristics such as control loop periodicity, subsystem activity, and operational state transitions.

### 3.2. Hardware Setup

The experimental setup that was used was comprised of an embedded CAN node and an external test and logging interface. The embedded system is based on a QT PY ESP32-S2 microcontroller, which interfaces with the CAN network through an MCP2515 SPI-based CAN controller. Communication between the ESP32-S2 and the MCP2515 is handled over the SPI bus, while a standard DB9 connector provides physical access to the CAN network.

CAN traffic is monitored and recorded using an external test system. Data capture is performed on a Dell Precision 5530 laptop running Ubuntu 20.04, equipped with an Intel Core i9 (8th generation) processor and 32 GB of RAM. A USB-to-CAN analyzer (APGDT002) connects the laptop to the CAN bus and is used to observe and log CAN messages in real time for subsequent analysis.

### 3.3. Software Implementation

This section describes the embedded and host-side software used to generate, transmit, and record encrypted CAN traffic during experimentation. First, the embedded device software is discussed. Then, the software used on the laptop is presented.

#### 3.3.1. Embedded System Software

The embedded software is designed to transmit CAN messages at a fixed rate of 100 Hz. This is a typical rate for real-time controls and sensor signals commonly found in automotive systems. The implementation runs on the ESP32-S2 microcontroller. It utilizes the SpeckSmall encryption library [11] for payload encryption and the MCP_CAN library [24] for CAN communication.

Two types of CAN messages are generated. The primary message type carries encrypted payload data. A secondary diagnostic message is transmitted in plaintext to support performance measurement and validation. The encrypted payload consists of a single-byte freshness value followed by application data, a constant reference value, and a mirrored data value. The freshness value is randomly generated for each transmission and ensures that identical application data is unlikely to produce repeated ciphertext, thereby offering the potential to limit replay and pattern-based analysis.

The diagnostic message mirrors the contents of the encrypted payload. Additionally, it includes timing measurements obtained using the ESP32-S2 micros() function. These measurements capture the time required for freshness value generation and payload encryption. This allows the computational overhead introduced by encryption to be evaluated independently of CAN transmission timing.

Message transmission timing is controlled using a hardware timer on the ESP32-S2. A timer interrupt sets a flag that triggers message generation and transmission at the desired 100 Hz rate, producing deterministic behavior consistent with real-time embedded control systems.

The application data field is implemented as a periodically increasing and decreasing signal, simulating a dynamic control or sensor value. The mirrored data field is deterministically related to the application data and is a correlated signal for evaluating the impact of encryption on inter-signal relationships.

### 3.3.2. Laptop Test and Recording Software

On the laptop, CAN traffic is captured and logged using a custom data-collection application built using the Robot Operating System (ROS). The software subscribes to incoming CAN frames and records message data to CSV files for offline analysis.

For each observed arbitration ID, a corresponding CSV file is created to organize traffic by message type. Each received CAN frame is logged with its data payload, data length code (DLC), and timestamp. This provides a temporal record of bus activity. This structure enables subsequent analysis of timing behavior, payload variability, and signal correlation using external analysis tools, such as Python.

This logging framework provides the data collection which is required for the experimental analysis presented in Section 4.

### 3.4. Research Questions

This study evaluates the feasibility and system-level impact of applying lightweight block cipher encryption to a high-frequency automotive control signal transmitted at 100 Hz. The experimental evaluation is structured to address several key research questions.

The first relates to the timing impact. This work seeks to determine whether the integration of Speck cipher encryption affects the inter-message timing required for real-time embedded control. The second relates to data masking. This study evaluates whether encryption is effective at obscuring constant-value transmissions on the CAN bus.

The third area of research is regarding pattern obfuscation. This work characterizes the extent to which encryption disrupts the incremental and decremental patterns in message payloads. Another key area is correlation disruption. This study evaluates whether encryption can prevent the detection of linear correlations between related CAN signals.

Next, measurement decorrelation is assessed. This work analyzes whether encryption reduces the observable correlation between application data on the CAN bus and externally measured signals. Finally, computational overhead is assessed to determine the computational timing cost introduced by freshness value generation and payload encryption.

Each research question is evaluated through targeted experiments described in Section 4.

## 4. Results

This section presents the experimentation and results for the research areas that were described in Section 3. Data was collected over a three-minute interval for both unencrypted and encrypted CAN traffic. This resulted in approximately 19,710 messages per configuration. Message timing behavior was evaluated by computing the time interval between consecutive CAN transmissions and analyzing the resulting distributions.

### *4.1. Timing Impact of Speck Cipher Implementation*

First, evaluation of whether payload encryption using the Speck cipher affects the timing consistency of CAN message transmissions in a 100 Hz control signal was conducted. Figure 4 shows the kernel density estimate of the inter-message intervals observed during encrypted message transmission.

Figure 4 shows that the measured time intervals are tightly clustered around 0.01 seconds (10 ms). This corresponds to the expected transmission period for a 100 Hz signal. The distribution exhibits minimal spread and no secondary modes. This indicates stable and consistent timing behavior.

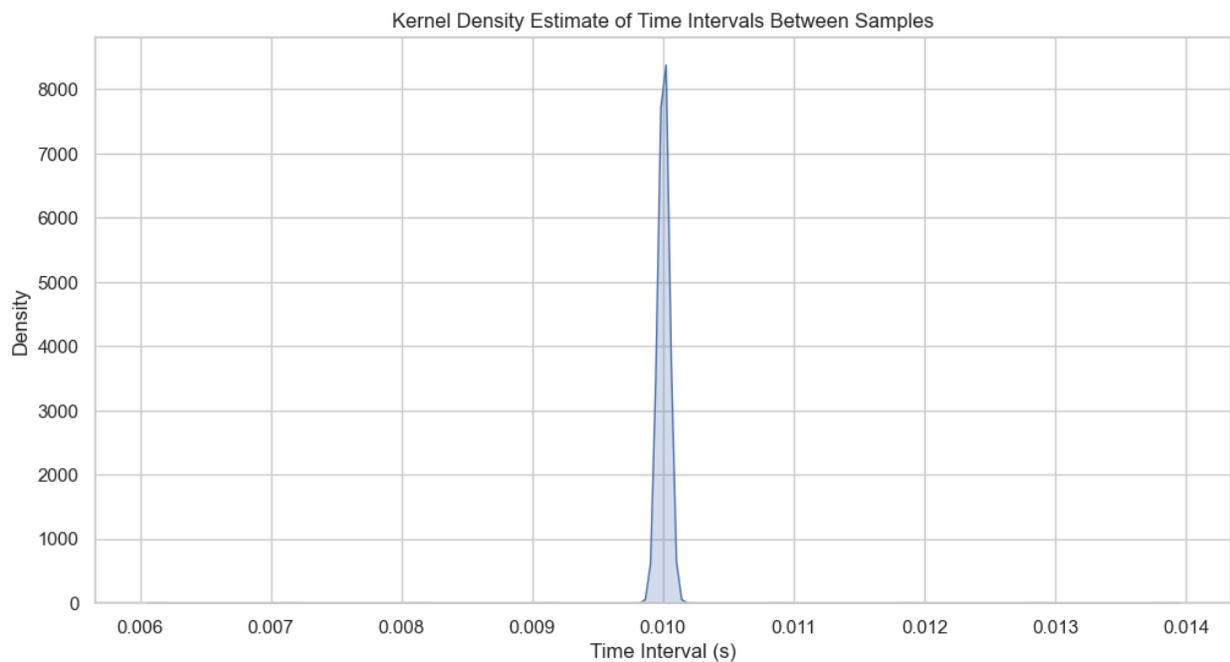

**Figure 4.** Calculated Time Intervals Between CAN Messages.

These results show that the inclusion of lightweight payload encryption does not introduce observable disruption to CAN message transmission timing at the system level. The encryption implementation preserves the intended 100 Hz communication rate, suggesting that the computational overhead of encryption does not interfere with real-time message scheduling.

### 4.2. Effectiveness of Encryption in Masking Constant Value Transmissions

Next, assessment turns to whether payload encryption effectively obscures constant-value fields in CAN messages, as these are a common target for semantic reverse engineering. A single payload byte (byte_2) was configured to transmit a constant value through both the encrypted and the unencrypted message streams. This allows a direct comparison of their behavior.

Figures 5 and 6 show the time-series values of byte_2 for encrypted and non-encrypted transmissions. In the non-encrypted case (shown on the right, in the figure), the byte value remains constant over time, showing the static nature of the underlying signal. This behavior would be readily identifiable through passive observation of CAN traffic.

In contrast, the encrypted representation of byte_2 (shown on the left side of Figures 5 and 6) varies across transmissions, despite the underlying application value remaining constant. It produces no visible pattern in the time series data. The zoomed view in Figure 6 shows that this variability exists at small time scales and is not an artifact of plot resolution.

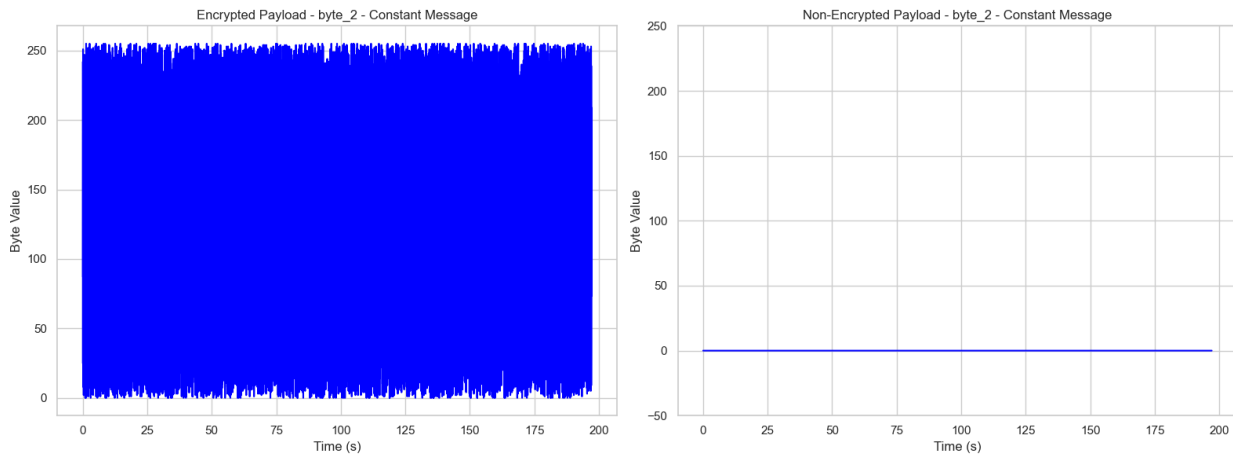

**Figure 5.** Constant value analysis – encrypted (left) and non-encrypted (right).

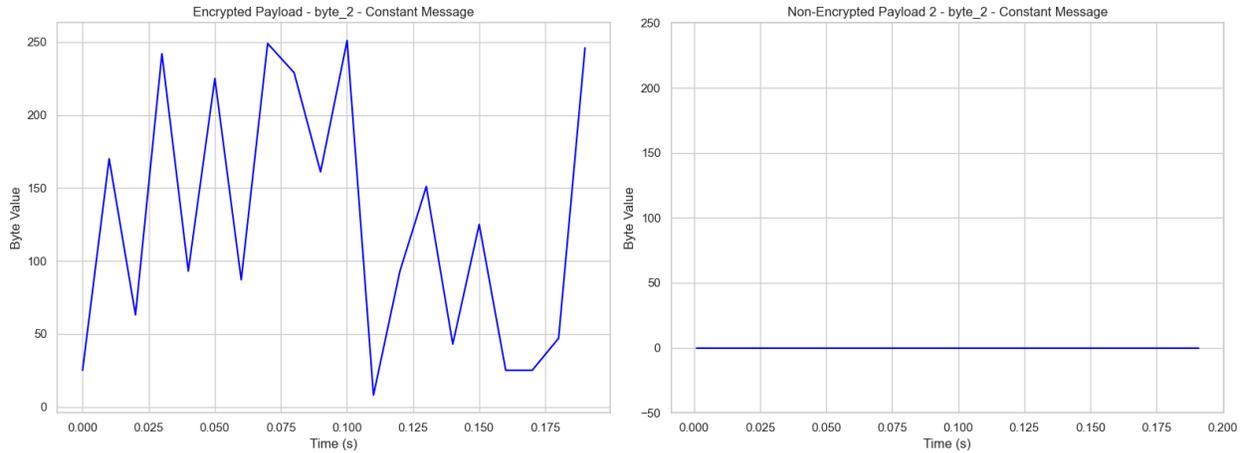

**Figure 6.** Constant value analysis – encrypted (left) and non-encrypted, zoomed in (right).

These results demonstrate that the lightweight payload encryption effectively masks constant-value CAN signals, eliminating a key feature commonly exploited during semantic taxonomy-based reverse engineering. By removing observable invariance in payload data, encryption reduces the ability of an attacker to infer signal semantics through passive traffic analysis.

## 4.3. Obscuring Incremental and Decremental Patterns

Now focus turns to evaluating the effectiveness of payload encryption in obscuring the incremental and decremental patterns that are commonly present in CAN application data. A single payload byte (byte_1), which is designated as application data (*appData*), was configured to cycle incrementally from 0 to 125 and then decrement back to 0 with each CAN message transmission.

Figures 7 and 8 show the time-series behavior of *appData* for both encrypted and non-encrypted transmissions. In the non-encrypted case (shown on the right in the figures), the incremental and decremental structure of the signal is clearly visible. A repeating sawtooth pattern, that would be readily identifiable through passive traffic observation, is produced.

In contrast, the encrypted representation of *appData* (shown on the left in the figures) exhibits no observable monotonic or cyclic structure. Although the underlying application signal follows a deterministic pattern, its encrypted form does not exhibit a visually discernible incremental or decremental behavior. The zoomed view in Figure 8 confirms that this lack of structure persists at smaller time scales.

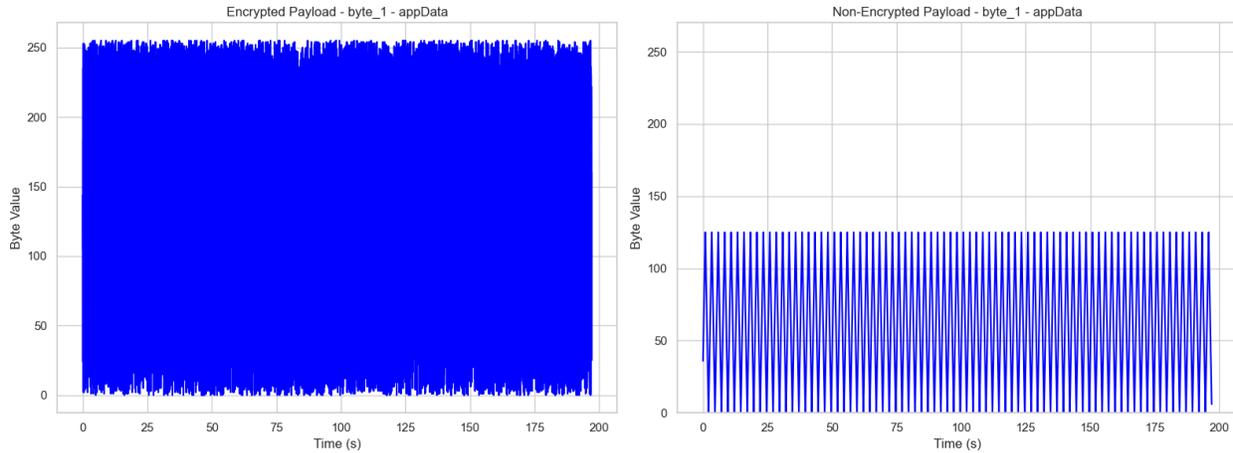

**Figure 7.** Application data analysis – encrypted (left) and non-encrypted (right).

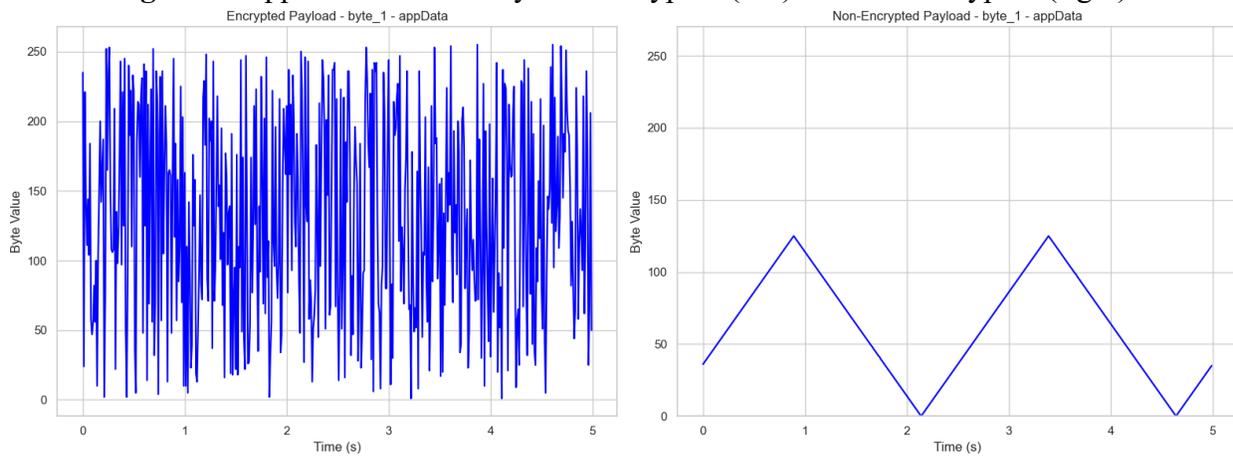

**Figure 8.** Application data analysis – encrypted (left) and non-encrypted, zoomed in (right).

These results demonstrate that lightweight payload encryption effectively masks predictable value progression in CAN messages. By eliminating observable trends in the application data, encryption reduces the ability of an attacker to infer signal semantics through pattern-based analysis. This is a key step in semantic taxonomy-based reverse engineering.

*4.4. Disruption of Observable Linear Correlations Between Related CAN Signals*

Next, the effectiveness of payload encryption in disrupting linear correlations between related CAN signals is evaluated. Two payload bytes were selected for analysis: the application data byte (*appData*, byte_1) and a deterministically related mirrored value (*shiftedData*, byte_3). These signals were constructed to exhibit a strong linear relationship, representative of the correlated measurements commonly present in automotive control systems.

Figures 9 and 10 show the time-series behavior of *appData* and *shiftedData* for non-encrypted transmissions. In the non-encrypted case, the relationship between the two signals is clearly observable, with both signals exhibiting synchronized, linearly related patterns over time. This deterministic relationship would be readily identifiable through passive CAN traffic observation and could potentially be exploited during semantic taxonomy-based reverse engineering.

In contrast, the encrypted representations of *appData* and *shiftedData*, which are shown in Figures 11 and 12, exhibit no visually discernible relationship. Although the underlying signals remain deterministically related, prior to encryption, their encrypted forms do not preserve an observable linear dependency. Quantitatively, the Pearson correlation coefficient drops from near unity, in the non-encrypted case, to approximately 0.016 after encryption. The zoomed in view in Figure 12 confirms that the lack of correlation is present at smaller time scales.

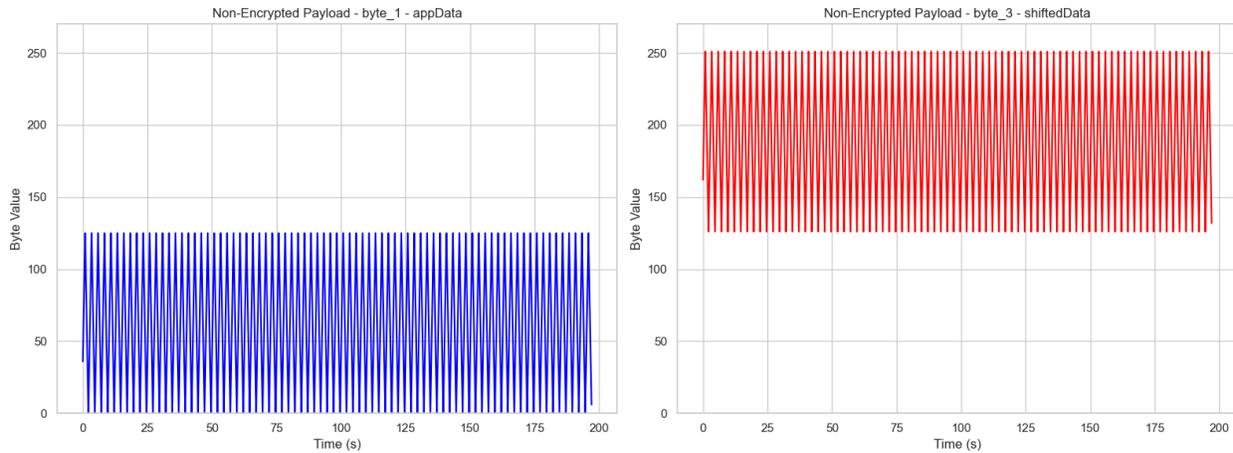

**Figure 9.** Application data (left) and mirrored data (right).

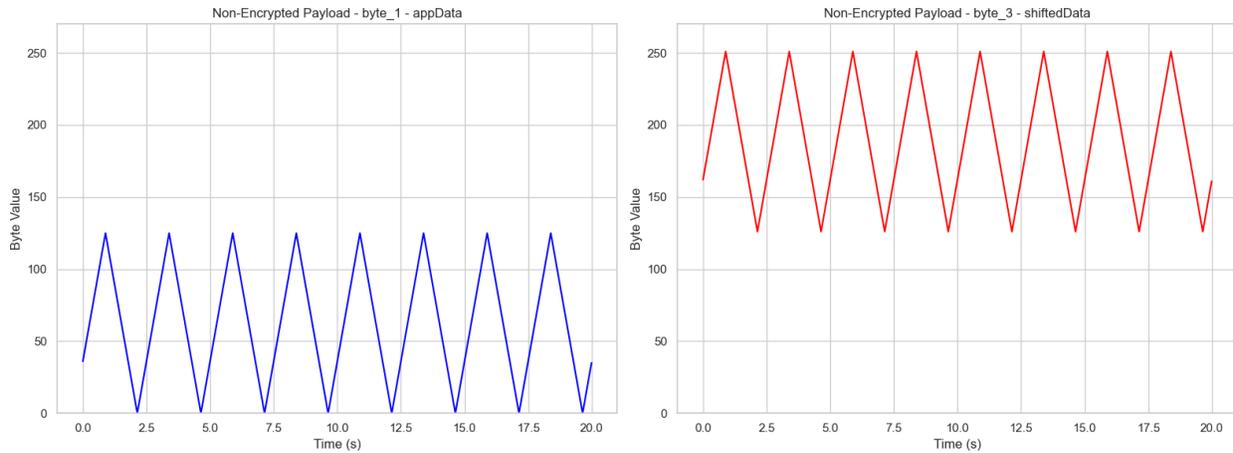

**Figure 10.** Application data (left) and mirrored data, zoomed in (right).

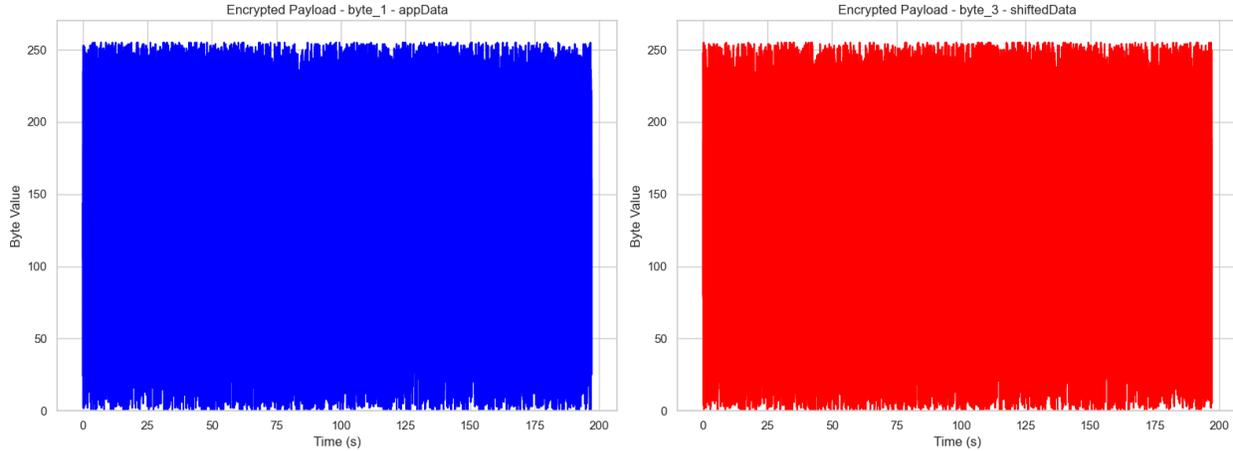

**Figure 11.** Encrypted application data (left) and mirrored data (right).

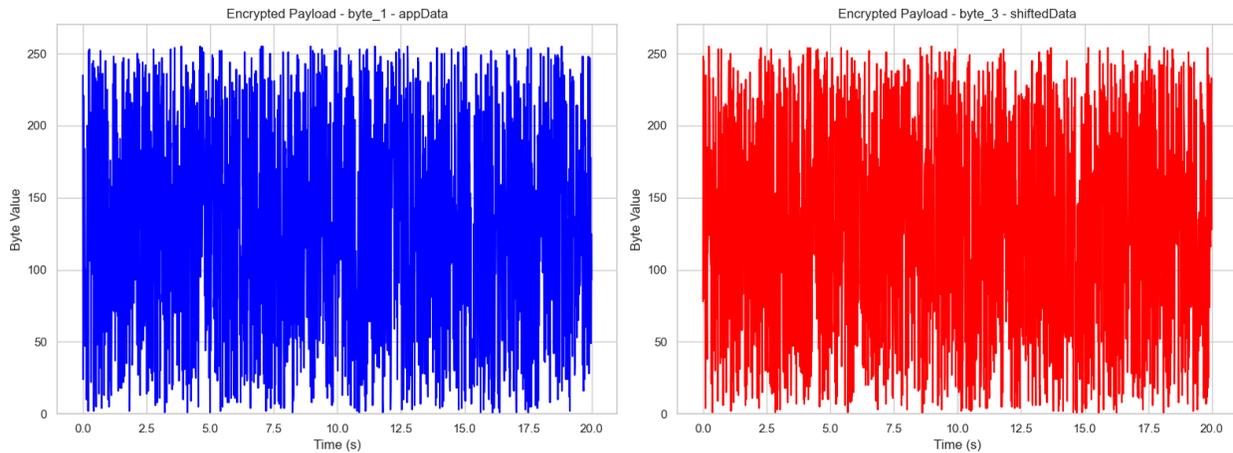

**Figure 12.** Encrypted application data (left) and mirrored data, zoomed in (right).

These results demonstrate that lightweight payload encryption effectively disrupts the observable linear relationships between the related CAN signals. By eliminating the observable inter-signal correlation, the encryption reduces the efficacy of reverse engineering techniques that rely on identifying deterministic relationships across message fields, further impeding semantic inference from CAN traffic.

### *4.5. Unlinking Data Correlations to Application Data on the CAN Bus*

The next focus is to evaluate the effectiveness of payload encryption in disrupting observable correlations between externally measured system signals and application data that are transmitted on the CAN bus. Two signals were considered: application data (*appData*, byte_1), transmitted on the CAN bus, and a corresponding measured value (*shiftedData*, byte_3), that represents a physical measurement that is linearly correlated to the application data, prior to encryption. Such correlations are commonly exploited during semantic taxonomy-based reverse engineering, when physical observations are used to help infer CAN signal semantics.

Figures 13 and 14 show the time-series behavior of the measured signal (*shiftedData*) alongside the encrypted representation of *appData*. Although the underlying application data remains deterministically related to the measured signal prior to encryption, this relationship is no longer observable, once encryption is applied to the CAN payload.

In the encrypted case, the application data exhibits no visually discernible relationship to the measured signal. Quantitatively, the Pearson correlation coefficient between the measured signal and the encrypted CAN data was calculated to be approximately -0.005, indicating the effective elimination of observable linear correlation. The zoomed view, shown in Figure 14, confirms that this lack of correlation persists at smaller time scales.

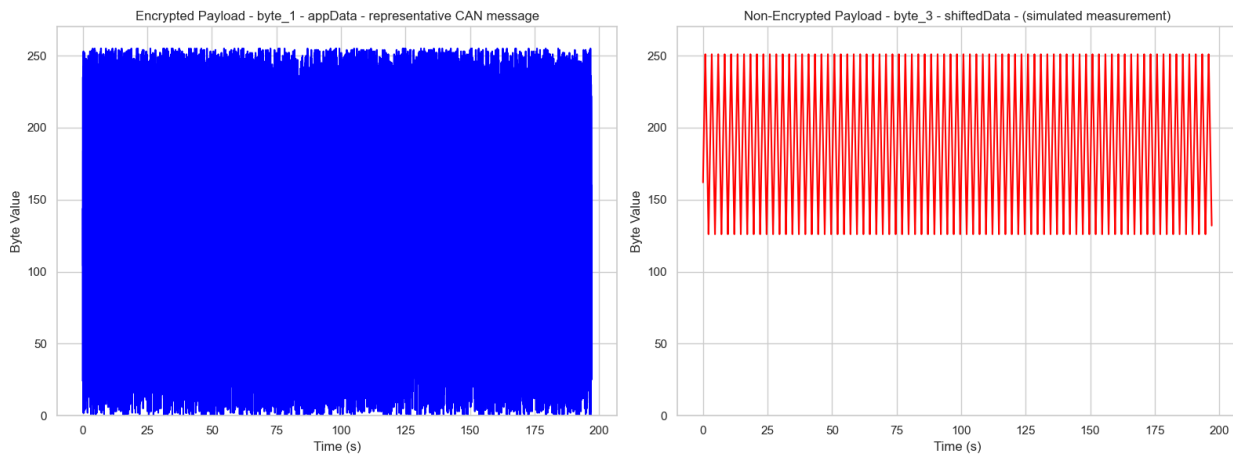

**Figure 13.** Encrypted CAN data (left) and known linear correlated measured values (right).

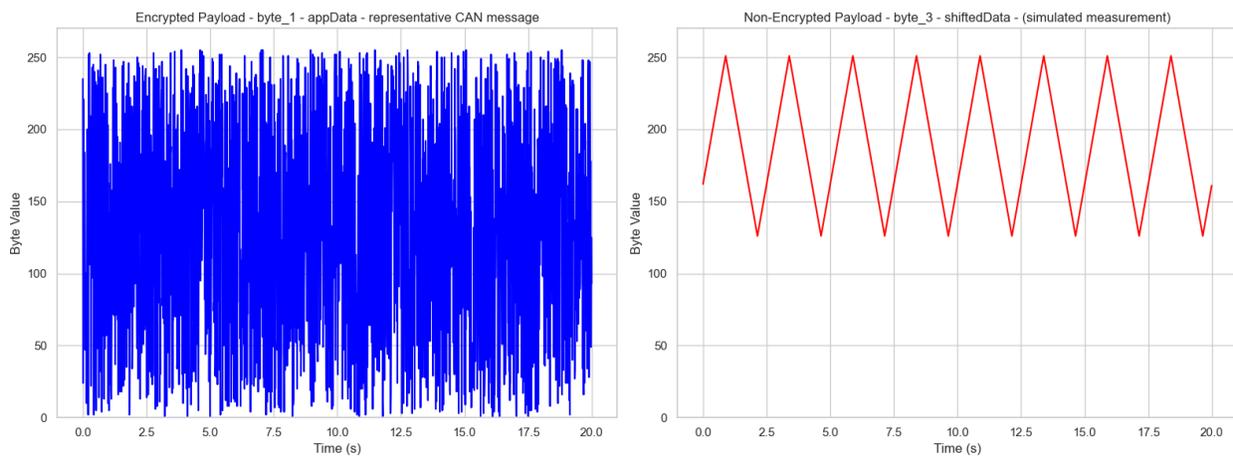

**Figure 14.** Encrypted CAN data (left) and known linear correlated measured values (right), zoomed in.

The data shows that lightweight payload encryption effectively severs observable correlations between physical measurements and the CAN application data. By preventing external measurements from being linked to encrypted CAN signals, encryption reduces the feasibility of approaches that rely on this correlation.

## 4.6. Overall Impact on System Timing

Now, the timing overhead introduced by payload encryption within the embedded CAN node is evaluated. Specifically, the execution time required for random freshness value generation and for encrypting the CAN payload is measured. These operations occur within the real-time execution path and therefore directly influence the feasibility of deploying encryption in time-critical automotive control systems.

Figure 15 shows the kernel density estimates of execution time for random freshness value generation and CAN payload encryption. The left plot illustrates the distribution of processing times required to generate a random freshness value, while the right plot shows the distribution of encryption durations for the CAN payload.

The freshness value generation process exhibits a narrow timing distribution centered at approximately 3 microseconds, as shown in figure 15. The small cluster of observations near 2 microseconds does not indicate instability in the freshness generation process. Instead, it is due to timer resolution granularity and instruction-level execution alignment within the microcontroller. Because the ESP32-S2 timing function provides microsecond-level resolution, minor variations in instruction scheduling or interrupt timing can cause measurements to register at either 2 or 3 microseconds. Importantly, this bounded variation remains deterministic and does not affect system-level timing behavior. The encryption routine demonstrates a narrow execution-time distribution centered near 20 microseconds. This limited dispersion suggests low temporal jitter and highly deterministic behavior, which is desirable in real-time embedded control systems.

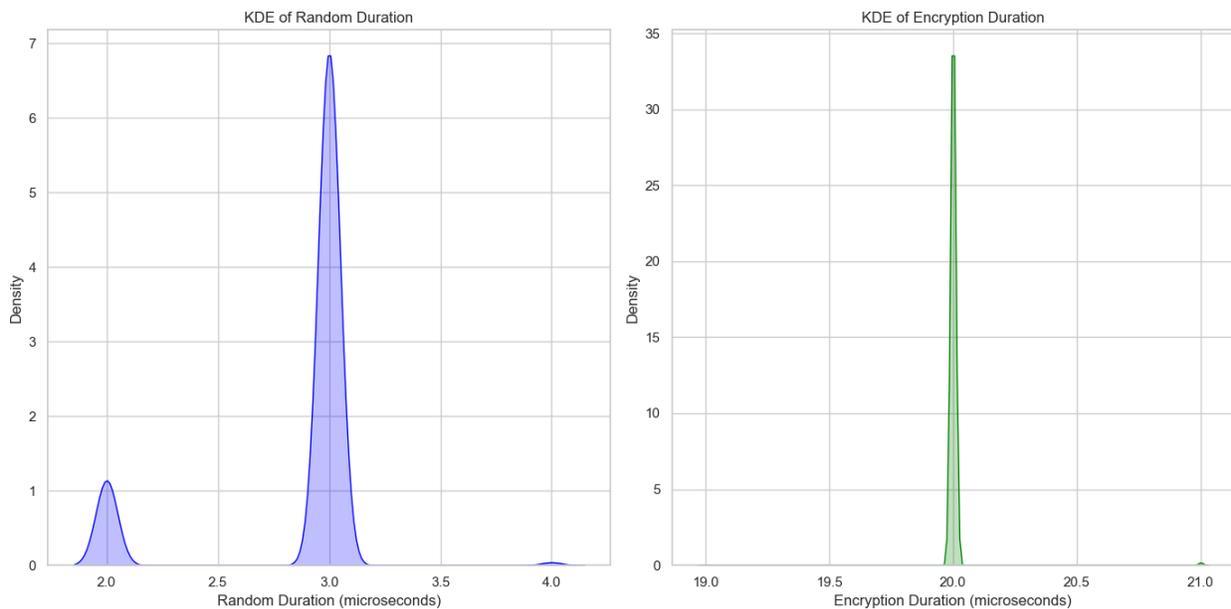

**Figure 15.** Timing for random number generation (left) and CAN payload encryption (right).

The combined execution time of approximately 23 microseconds remains below the 10-millisecond interval required for 100 Hz CAN message transmission. Even accounting for microsecond-scale variability, the measured overhead represents less than 0.25% of the message period. Therefore, the evaluated encryption strategy does not introduce timing behavior that would violate real-time constraints, supporting its feasibility for high-frequency CAN communication

## 5. Discussion

This section interprets the experimental results presented in Section 4 in the context of semantic taxonomy-based reverse engineering of CAN traffic. Each experiment addressed a distinct aspect of signal observability, timing feasibility, or correlation inference. Collectively, they facilitate evaluation of whether lightweight payload encryption can meaningfully increase resistance to passive analysis, without violating real-time constraints.

The first experiment demonstrated that lightweight payload encryption does not introduce observable disruption to CAN message timing at a 100 Hz transmission rate. The preservation of a tightly clustered inter-message interval indicates that the encryption overhead does not interfere with message scheduling or periodicity, under the tested conditions. This result is important, because any mechanism that alters timing behavior could become a distinguishing feature or violate real-time control requirements.

The results of the second experiment show that payload encryption eliminates observable constant-value fields in CAN messages. Constant signals are among the easiest features to identify during passive traffic analysis and often serve as anchors for semantic labeling. By removing observable invariance in payload data, encryption directly impedes one of the most common entry points for semantic reverse engineering.

The third experiment extended this analysis to dynamic signals exhibiting monotonic structure. Incremental and decremental patterns are frequently associated with sensor readings or control commands and can be readily identifiable in unencrypted CAN traffic. The absence of observable monotonic structures in the encrypted payloads indicates that encryption effectively suppressed these semantic cues. This further limits an attacker's ability to infer signal meaning from value progression.

The fourth experiment demonstrated that encryption disrupts observable linear relationships between related CAN signals. Such correlations are commonly exploited to infer functional relationships between signals and to validate semantic hypotheses. By eliminating the observable inter-signal correlation, encryption increases the effort required to reconstruct system behavior through the statistical analysis of CAN traffic.

The fifth experiment addressed a more advanced reverse engineering scenario where external physical measurements are correlated with CAN message content. The results of this experiment indicate that encrypting CAN payloads reduces the observability of the relationship between external measurements and transmitted application data. This limits the effectiveness of combined physical and network-based observation, which can be used to accelerate semantic inference in cyber-physical systems.

Finally, the sixth experiment confirmed that the computational overhead associated with freshness value generation and payload encryption remains limited and consistent. The measured processing times are well below the timing margin for a 100 Hz control signal, indicating that encryption does not introduce timing behavior that is incompatible with real-time embedded operations. This result suggests the practical feasibility of deploying lightweight encryption on resource-constrained CAN nodes.

## 6. Conclusion

This study has evaluated the feasibility of applying lightweight payload encryption to CAN communications using resource-constrained embedded hardware. By implementing the SpeckSmall encryption library on a QT PY ESP32-S2 microcontroller, this study has examined whether encryption can meaningfully reduce the observability of CAN message content without violating real-time communication constraints.

The experimental results have shown that payload encryption does not disrupt CAN message timing at the 100 Hz transmission rate, preserving the deterministic behavior required by real-time automotive control systems. Additionally, encryption effectively obscures constant-value fields, suppresses predictable value progression, and disrupts observable correlations between related CAN signals. These effects collectively reduce the feasibility of semantic taxonomy-based reverse engineering through passive traffic analysis.

These findings indicate that lightweight encryption can be integrated into embedded CAN nodes with limited computational overhead. This suggests its practicality for deployment in resource-constrained environments that are typical of automotive systems. This study provided a proof-of-concept implementation and did not address all aspects of cryptographic security. It also provided empirical evidence that encryption reduces observable structures that can be used for signal inference without compromising system timing behavior.

Future work could potentially explore the integration of authentication mechanisms, expanded threat models. It could also evaluate the proposed approach with other automotive and industrial communication protocols. Validation in diverse operational environments could provide deeper insight into system resilience under more complex attack scenarios.

Overall, this study has provided empirical evidence that lightweight payload encryption can reduce passive observability of CAN signal semantics without violating timing behavior under the tested conditions. This suggests that future study in this area is merited.